\documentclass{article}

\usepackage{arxiv}

\usepackage[utf8]{inputenc} 
\usepackage[T1]{fontenc}    
\usepackage{hyperref}       
\usepackage{url}            
\usepackage{booktabs}       
\usepackage{amsfonts}       
\usepackage{nicefrac}       
\usepackage{microtype}      
\usepackage{multirow}
\usepackage{amsmath,amssymb}
\usepackage{graphicx}
\title{On the modeling of thermal and free carrier nonlinearities in Silicon On Insulator microring resonators}

\author{
  Massimo Borghi\thanks{corresponding author: massimo.borghi@unitn.it},  Davide Bazzanella, Mattia Mancinelli and Lorenzo Pavesi \\
  Nanoscience laboratory, Department of Physics\\
  University of Trento\\
  Via Sommarive 14, 38123, Trento, Italy
}

\begin{document}
\maketitle

\begin{abstract}
The temporal dynamics of integrated silicon resonators has been modeled using a set of equations coupling the internal energy, the temperature and the free carrier population. Owing to its simplicity, Newton's law of cooling is the traditional choice for describing the thermal evolution of such systems. In this work, we theoretically and experimentally prove that this can be inadequate in monolithic planar devices, leading to inaccurate predictions. A new equation, that we train to reproduce the correct temperature behaviour, is introduced to fix the discrepancies with the experimental results. We discuss the limitations and the range of validity of our refined model, identifying those cases where Netwon's law provides, nevertheless, accurate solutions. Our modeling describes the phenomena underlying thermal and free carrier instabilities, and is a valuable tool for the engineering of photonic systems which relay on resonator dynamical states, such as all optical spiking neural networks or reservoirs for neuromorphic computing.   
\end{abstract}


\section{Introduction}
\label{section_1}
Microresonators are now ubiquitous in many fields of integrated photonics, ranging from optical communications \cite{testa2018integrated}, bio-sensing \cite{steglich2019optical}, spectroscopy \cite{suh2016microresonator}, frequency metrology \cite{papp2014microresonator} and quantum optics \cite{llewellyn2020chip}. Silicon microresonators have been historically appealing due to their compact foot-print, large third order nonlinearities and wide bandwidth of operation \cite{borghi2017nonlinear}. A bottleneck when they are operated in the C-band is the presence of Two Photon Absorption (TPA) and Free Carrier Absorption (FCA), which introduce additional losses to the device \cite{borselli2005beyond}. Irrespective of the application, these have often a detrimental role; examples are the limitation of the maximum attainable Signal to Noise Ratio (SNR) in optical links \cite{de2019power}, gain saturation in wavelength conversion devices \cite{wang2016influence} and reduction of the heralding efficiency in parametric photon pair sources \cite{sinclair2019temperature}. Nonetheless, there are some exceptions in which TPA and FCA in resonators are sought for all optical information processing, such as memory storage \cite{almeida2004optical}, logic operations \cite{xu2007all} or unidirectional propagation \cite{fan2012all}. TPA creates electron-hole pairs whose relaxation self-heats the resonator, so that the temperature and the carrier dynamics becomes coupled \cite{johnson2006self}. Their interplay may appear at the output of the resonator as time-instabilities associated to Self-Pulsing (SP), bistability, excitability, or as a nonlinear distortion in the transmitted spectrum \cite{de2019power,luo2012power,zhang2014experimental,zhang2013multibistability,takemura2020designs}. There is a growing interest in exploiting these regimes for neuromorphic photonics, making silicon resonators promising candidates for the realization of spiking neural networks, single perceptrons, or extended reservoirs \cite{li2019photonic,catuneanu2019nonlinear,denis2018all,mesaritakis2013micro,van2012cascadable}. The design and optimization of these systems require the accurate modeling of each node, which is traditionally implemented using a set of three coupled ordinary differential equations (ODE): one for the internal energy of the resonator, one for the temperature and one for the free carrier population \cite{van2012simplified}. The linearization of this set around equilibrium points, combined with the tool of Hopf bifurcations, allows to identify the optimal regions of operation in the parameter space defined by the input power $P$ and the resonance detuning $\Delta\nu$ of the pump laser \cite{van2012simplified,van2012cascadable,zhang2013multibistability}.  
In this work, we experimentally and theoretically show that there are cases in which the commonly used ODEs fail to describe the dynamics of the silicon microresonator. This happens whenever the Newton's law of cooling, from which the temperature equation derives, does not  approximate the actual thermal dynamics. Using Finite Element Method (FEM) simulations, we show that in our planar geometry of a silicon racetrack resonator embedded in a silicon dioxide cladding, the temperature is described by a fast ($\sim 100\,\textrm{ns}$) and a slow ($\sim 1\,\mu\textrm{s}$) relaxation time. The latter can be observed when self oscillations of similar period are internally established due the interplay between thermal and free carrier nonlinearities. In our case, a SP regime of sub-MHz frequency establishes as a consequence of a relatively long ($\sim 40\,\textrm{ns}$) free carrier lifetime.  We introduce a new temperature equation that reproduces such temporal features, leading to predictions very close to the experimental observations. We investigate the generalization capabilities of our model to other resonators with similar geometry but different quality factors, proposing and experimentally validating a simple method to assess the error on the predicted SP period. Finally, we discuss the conditions where Newton's law can be applied, and the artifacts which arise on the specific heat, on the carrier lifetime or on the carrier dispersion coefficient when an inadequate model for temperature is used to fit the experimental data.

\section{Observation of sub-MHz self pulsing}
\label{section_2}
Our racetrack microresonators (MR) have been fabricated at the IMEC/Europractice facility within a  multi-project wafer run on a $200\,\textrm{mm}$ SOI wafer of $220\,\textrm{nm}$ thickness. Silicon waveguides are embedded in a silica cladding, and have a width of $450\,\textrm{nm}$ to ensure single mode operation on both polarizations. 
\begin{figure}[h!]
\centering\includegraphics[scale = 0.92]{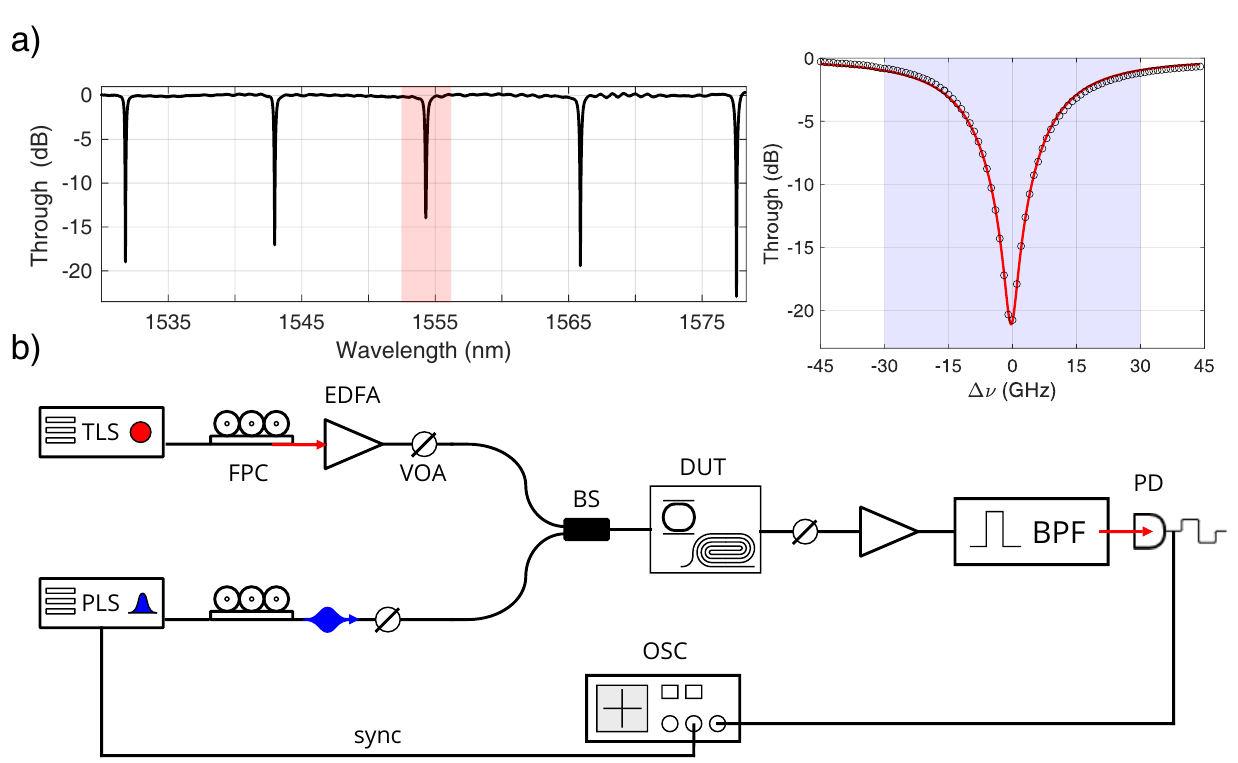}
\caption{(a) Low power transmission spectra of the MR, collected at the Through port. Highlighted in red is the resonance order where the SP regime is analyzed. An enlarged view of this resonance is shown on the right panel. Here, the blue region marks the range of laser detuning $\Delta\nu$ investigated in this work. (b) The experimental setup implemented to measure the SP of the MR and the carrier lifetime. \mbox{TLS = Tunable Laser Source}, \mbox{PLS = Pulsed Laser Source}, \mbox{FPC = Fiber Polarization Controller}, \mbox{EDFA = Erbium Doped Optical Amplifier}, \mbox{VOA = Variable Optical Attenuator}, \mbox{BS = Beam Splitter}, \mbox{DUT = Device Under Test}, \mbox{BPF = Band Pass Filter}, \mbox{PD = Photodiode}, \mbox{OSC = Oscilloscope}.\label{fig1}}
\end{figure}
An Add-Drop filter configuration is realized by evanescently coupling two bus waveguides to the MR of $7\,\mu\textrm{m}$ of radius. This is accomplished through a $250\,\textrm{nm}$ gap and a coupling length of $3\,\mu\textrm{m}$, which yields a measured coupling coefficient of $\kappa^{2}=0.063(3)$. From the low-power transmission spectra recorded at the Through port, shown in Fig.\ref{fig1}(a), we extracted an intrinsic quality factor (Q) of $Q_{\textup{i}} = 1.11(8)\times10^5$ and a loaded Q of $Q_{\textup{L}} = 6.5(2)\times10^{3}$.
We used the experimental setup shown in Fig.\ref{fig1}(b) to probe the resonator dynamics. A Continuous Wave (CW) tunable laser at $1555\,\textrm{nm}$ is amplified and injected at the input waveguide of the MR through a grating coupler ($\sim 3.5\,\textrm{dB}$ insertion loss). The input power is controlled by a Variable Optical Attenuator. Optionally, a C-band pulsed laser source of $40\,\textrm{ps}$ pulse width and $1\,\textrm{MHz}$ of repetition rate can be combined to the CW laser to perform pump and probe experiments. Light transmitted from the Drop port of the resonator is coupled off-chip by a second grating coupler, subsequently amplified and directed to a fast photodiode with $20\,\textrm{GHz}$ bandwidth.
\begin{figure}[t!]
\centering\includegraphics[scale = 0.62]{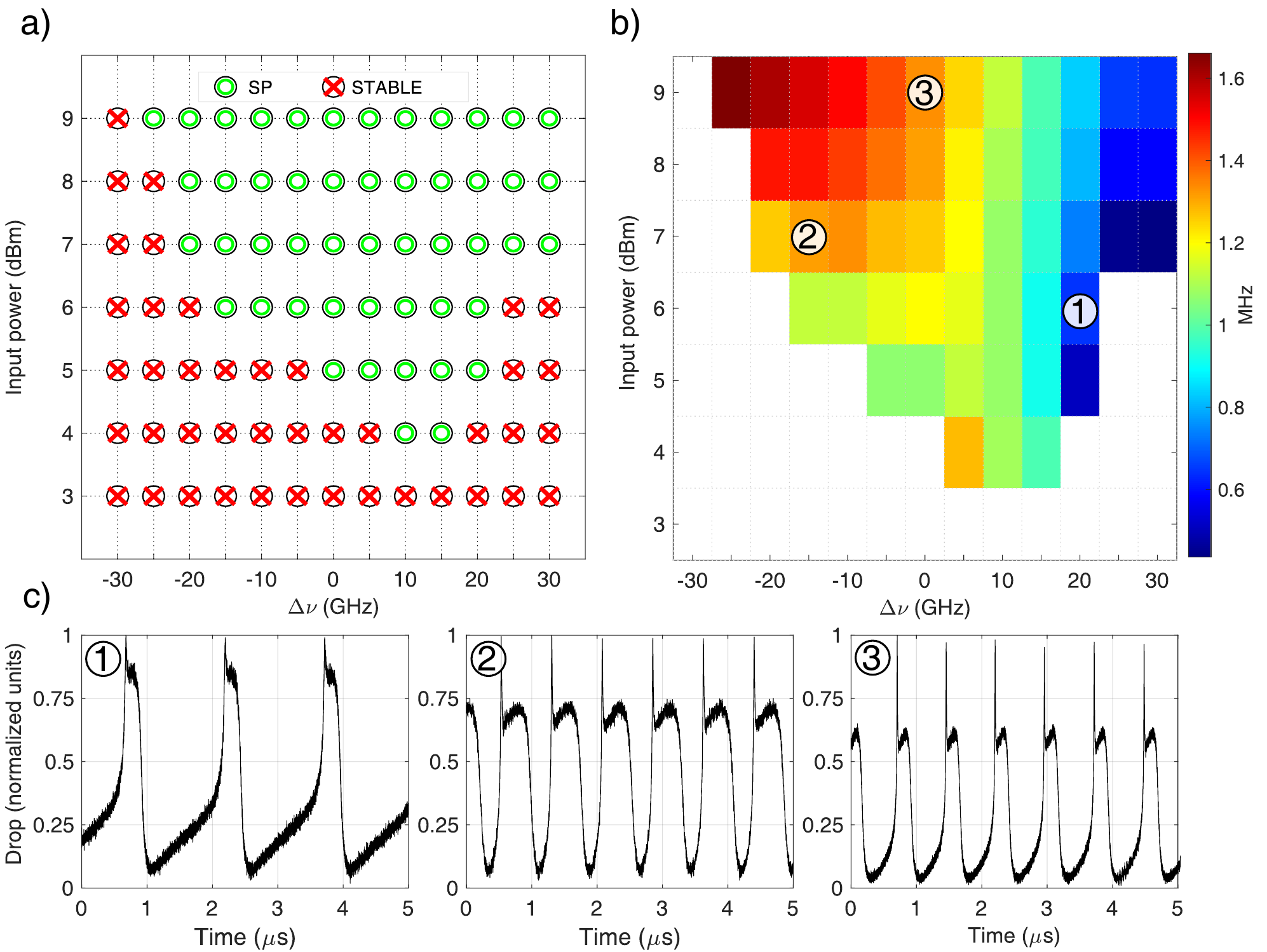}
\caption{(a) Stability map of the MR in the $\Delta\nu$ = laser frequency detuning, P = input power plane. Red crosses indicate the points where the MR, after an initial transient, shows stable output. Green circles indicate the points where the MR is SP. (b) Map of the oscillation frequency of the self pulsing regime. (c) Examples of time traces recorded at the output of the Drop port of the MR. The maximum of the intensity is normalized to one. The labels (1), (2) and (3) refer to the values of $(P,\Delta\nu)$ associated to each trace, which are indicated in panel (b). \label{fig_2}}
\end{figure}
A tunable band-pass filter centered at the laser wavelength is used to remove the spontaneous emission of the amplifiers hence increasing the SNR. A $40\,\textrm{GS}$ oscilloscope records the output of the photodiode.
\noindent As a first experiment, we investigated the stability regions of the MR in the two-dimensional plane $(P,\Delta\nu)$. The map shown in Fig.\ref{fig_2}(a) marks with circles the points where SP is observed, and with crosses the ones where the system has a stable CW output. This experiment is performed by resetting the initial conditions at each point $(P,\Delta\nu)$ by closing the input VOA shutter for few seconds. Self pulsing is observed at both positive and negative detunings, predominantly on the blue side of the resonace. Despite the similar MR geometry, this contrasts with the results in \cite{zhang2014experimental}, where SP is exclusively detected at the blue side of the resonace. According to the linear stability analysis in \cite{van2012simplified}, the blue shift of the SP region indicates that free carrier dispersion (FCD) is dominating over the thermo optic effect (TOE), which will find justification in Section \ref{section_3} and Section \ref{section_5}. The map shown in Fig.\ref{fig_2}(b) reports the frequency of SP, which we extract from time traces similar to those shown in Fig.\ref{fig_2}(c). The general trend agrees with the results of \cite{johnson2006self,zhang2014experimental}, where it is reported that the SP frequency increases from positive to negative detunings and by increasing the input power. The time traces in Fig.\ref{fig_2}(c) show all the distinctive features of SP induced by the interplay between FCD and TOE: the cycle starts with a free carrier bistability (narrow peak in the drop signal), which is then counteracted by a slower thermal shift pulling the resonance frequency in the opposite direction, partially restoring the initial transmittivity. When the thermal shift overwhelms FCD, the carrier concentration suddenly drops, leaving an hot cavity completely out of resonance (drop of the transmittivity in Fig.\ref{fig_2}(c)). This is followed by a slow thermal relaxation, where the temperature of the MR decays until the initial conditions are restored and a new cycle of oscillation begins.  
As shown in Fig.\ref{fig_2}(b), the average SP frequency is of the order of $\sim\,1\,\textrm{MHz}$, and can be as low as $\sim\,400\,\textrm{kHz}$ for large positive detunings. This is an order of magnitude lower than the SP frequencies reported in \cite{mancinelli2014chaotic,zhang2014experimental}, despite the very similar MR geometries. In general, SP periods of $\sim100\,\textrm{ns}$ arise when the thermal and the carrier lifetime are respectively of the order of $\tau_{\textup{th}} \sim 80-150\,\textrm{ns}$ and $\tau_{\textup{fc}}\sim 1-10\,\textrm{ns}$ \cite{van2012simplified}. This is because a complete cycle of oscillation requires both thermal and carrier relaxation, so its period is roughly given by $\tau_{\textup{fc}}+\tau_{\textup{th}}$. In order to justify a sub-MHz dynamics, a longer thermal and/or carrier lifetime has to be postulated. Since the first can be assessed from a FEM simulation, we performed a second experiment to measure the carrier lifetime.

\section{Determination of the free carrier lifetime}
\label{section_3}
\begin{figure}[h!]
\centering\includegraphics[scale = 0.65]{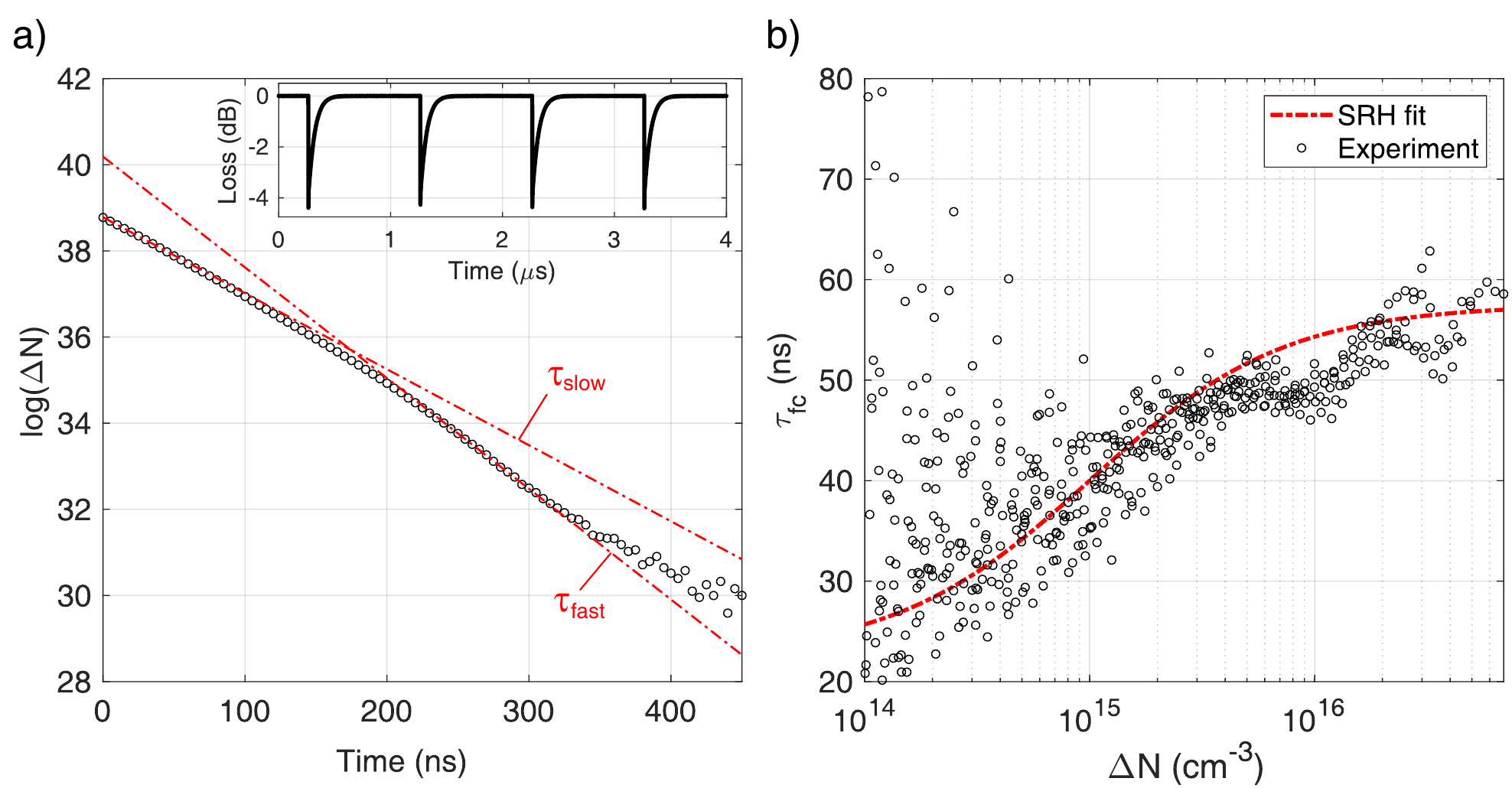}
\caption{(a) Temporal decay of the carrier concentration $\Delta N$ subsequent to the absorption of a pump pulse at $t=0$. The two red dashed lines indicate the expected profile for a purely exponential decay, i.e., $\log(\Delta N)=\log(\Delta N_0)-\frac{t}{\tau_{\textup{fc}}}$. The line indicated as $\tau_{\textup{slow}}\sim60\,\textrm{ns}$ is obtained by evaluating the average local slope for $0\leq t\leq50\,\textrm{ns}$, while the time interval $250\leq t\leq300\,\textrm{ns}$ is used to evaluate the slope of the line labeled as $\tau_{\textup{fast}}\sim25\,\textrm{ns}$. The inset shows the time-dependent transmittivity of the CW probe beam. Dips mark the time arrival of the pump pulses, which locally induce extra-losses due to free carrier absorption. The transmission has been normalized by taking into account the linear propagation loss. (b) Measured (black dots) carrier lifetime $\tau_{\textup{fc}}$ as a function of the carrier concentration. This is extracted starting from curves similar to the one in panel (a), and by using the method outlined in \cite{aldaya2017nonlinear}. The red dashed line is a fit of the lifetime obtained by implementing the Shockley Read Hall expression in Eq.(\ref{eq_1}) and the parameters listed in Table \ref{tab_1} of Appendix B. \label{fig_3}}
\end{figure}

The carrier lifetime is determined with a pump and probe technique applied to a spiral waveguide of length $7.1\,\textrm{mm}$, located on the same die of the MR. As shown in Fig.\ref{fig1}(b), a strong pulsed pump ($40\,\textrm{ps}$) with $\sim 1.5\,\textrm{W}$ peak power and a wavelength of $1550\,\textrm{nm}$ is combined with the CW probe and injected into the spiral. The high intensity of the pump creates an initial free carrier population $\Delta N_0$ through TPA, which recombines via trap states located at the Si/SiO$_2$ interface \cite{dimitropoulos2005lifetime}. During this transient regime, the CW probe experiences an extra-loss due to FCA, whose temporal profile follows the instantaneous free carrier population $\Delta N(t)$ along the waveguide. The time-dependent transmittivity of the spiral (see inset in Fig.\ref{fig_3}(a)) is used to extract the instantaneous carrier lifetime $\tau_{\textup{fc}}$, using the approach detailed in \cite{aldaya2017nonlinear}. Figure \ref{fig_3}(b) shows $\tau_{\textup{fc}}$ as a function of the carrier concentration $\Delta N$ inside the waveguide. 

\noindent The value of $\tau_{\textup{fc}}$ is bounded between $20\,\textrm{ns}$ and $ 60\,\textrm{ns}$. As shown in Fig.\ref{fig_3}(b), at the very early stage, carrier recombines with a characteristic lifetime of $\tau_{\textup{slow}}\sim 60\,\textrm{ns}$, which is significantly larger than the one in the late stage ($\tau_{\textup{fast}}\sim\,25 \textrm{ns}$), in which the carrier concentration has dropped by nearly two orders of magnitude. Even though the instantaneous lifetime is larger than the ones traditionally reported for SOI waveguides \cite{pernice2010time,almeida2004all,van2009silicon}, this is not entirely surprising. Indeed, as reported in \cite{aldaya2017nonlinear}, the free carrier lifetime appears to evolve nonlinearly with the carrier concentration, and ultra-long lifetimes of the order of $300\,\textrm{ns}$ have been observed. It is worth to mention that the waveguide geometry and the fabrication facility adopted in \cite{aldaya2017nonlinear} is the same of our work. \\
\noindent Trap-assisted recombination leads to electron-hole lifetimes which depend on the amount of excess carriers \cite{linnros1998carrier,shockley1952statistics,nomura1961decay}. When the density of traps is small compared to the equilibrium concentration of the majority carriers, the following Shockley Read Hall (SRH) expression holds \cite{shockley1952statistics}:
\begin{equation}
\tau_{\textup{fc}}(t) = \frac{\Delta N(t)}{R(t)} = \tau_0\frac{1+a\Delta N(t)}{1+b\Delta N(t)} \label{eq_1}
\end{equation}
where $R(t)$ is the carrier recombination rate, $\tau_0$ is the lifetime in the low carrier injection regime and $(a,b)$ are constants which depend on material properties, such as the doping level, the electron-hole capture cross-section and the trap energy level. The definition and the numerical values of these terms is reported in Table \ref{tab_1} of Appendix B. Equation (\ref{eq_1}) is used to model the points in Fig.\ref{fig_3}(b), indicating a good agreement with the experiment.
In principle, if the density of traps is of the same order of magnitude of the equilibrium concentration of the majority carriers, the phenomenon of trapping makes electrons and holes to recombine at different rates, so one has to separately describe the evolution of the two types of carrier \cite{aldaya2017nonlinear,nomura1961decay}. As a consequence, $\tau_{\textup{fc}}$ is not solely determined by the carrier concentration, but also by the initial excess carrier density. An experimental signature of the presence of trapping is the dependence of the carrier lifetime on the pump power that is used to create the initial excess population \cite{aldaya2017nonlinear}. Through numerical simulations, we verified that in the concentration range $10^{15}-10^{18}\,\textrm{cm}^{-3}$, which is of concern in our work, the phenomenon of trapping has a negligible impact on $\tau_{\textup{fc}}$.

\section{Evidences of inadequacy of the standard coupled equations}
\label{section_4}
\begin{figure}[t!]
\centering\includegraphics[scale = 0.65]{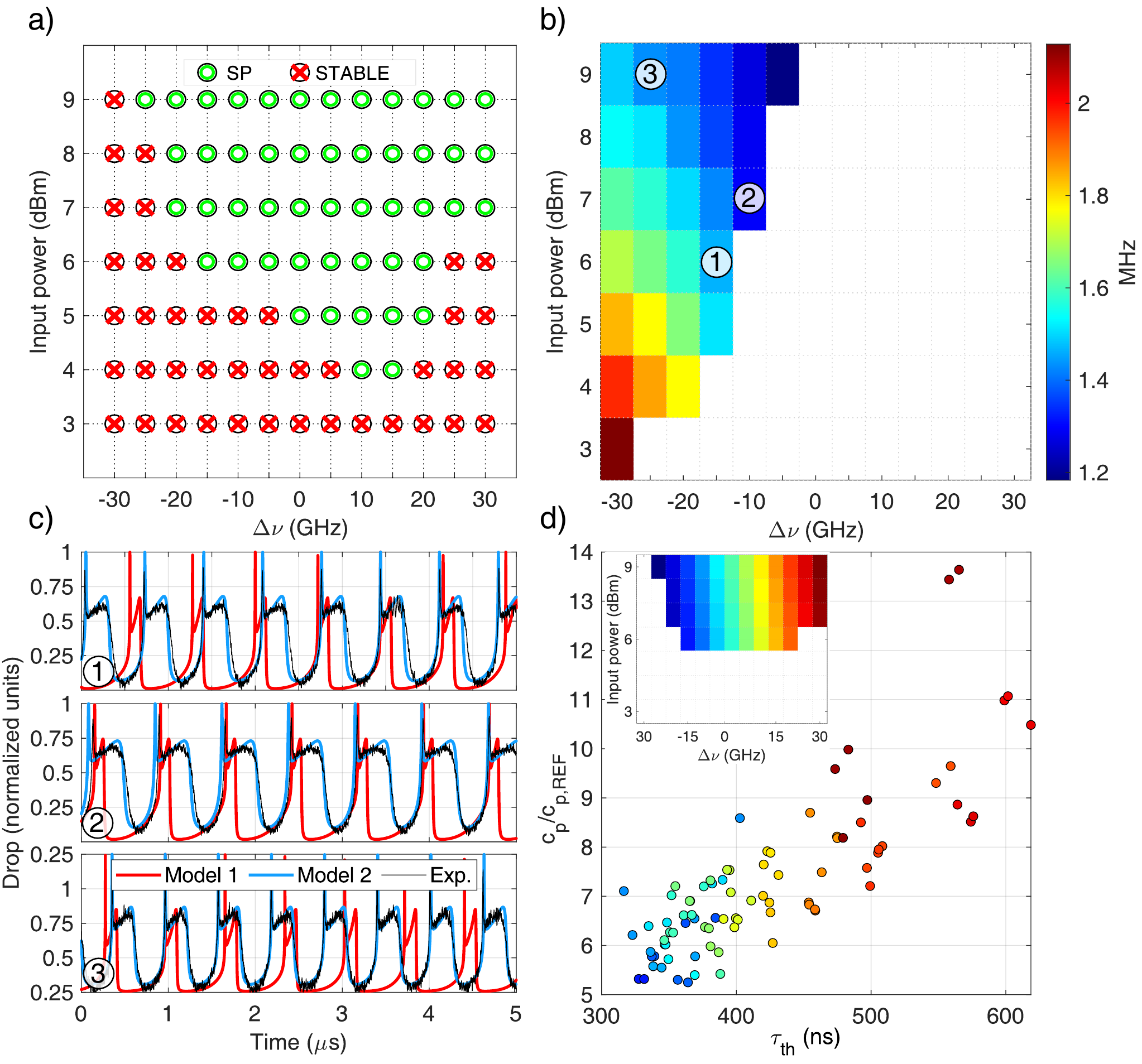}
\caption{(a) Simulated stability map obtained by integrating Eqs.(\ref{eq_2.1}-\ref{eq_2.3}) and by using the set of parameters listed in Table \ref{tab_2} of Appendix A. (b) Simulated frequency of the SP cycles. (c) Examples of experimental (black) and simulated time traces at the Drop port of the resonator, corresponding to points (1), (2) and (3) in the $(P,\Delta\nu)$ plane shown in panel (b). Red lines use Eqs.(\ref{eq_2.1}-\ref{eq_2.3}) with $c_p=c_{\textup{p,ref}} = 700\,\textrm{J/(KgK)}$ and $\tau_{\textup{th}}=275\,\textrm{ns}$. Blue lines use Eqs.(\ref{eq_2.1}-\ref{eq_2.3}) and the values of $c_p$ and $\tau_{\textup{th}}$ shown in panel (d). (d) Normalized specific heat $c_p/c_{\textup{p,ref}}$ as a function of the thermal lifetime $\tau_{\textup{th}}$. The color of the dots labels the position in the plane $(P,\Delta\nu)$. This color code is shown in the upper left inset. \label{fig_4}}
\end{figure}

The temporal dynamics of silicon MR is commonly modeled by a set of three, coupled ordinary differential equations, which govern the internal energy $U_{\textup{int}}=|a|^2$ of the resonator, the free carrier population $\Delta N$, and its differential temperature with the cladding $\Delta T$ \cite{johnson2006self}:
\begin{align}
\frac{da}{dt} & = i(\omega_0(1+\delta)+i\gamma))a+i\sqrt{\gamma_e}p_{\textup{in}} \label{eq_2.1}\\
\frac{d\Delta N}{dt} & = -\frac{\Delta N}{\tau_{\textup{fc}}}+g_{\textup{TPA}}|a|^4  \label{eq_2.2}\\
\frac{d\Delta T}{dt} & = -\frac{\Delta T}{\tau_{\textup{th}}}+\frac{P_{\textup{abs}}}{mc_p} \label{eq_2.3}\\
\delta & = -\frac{1}{n_0}\left( \left( \frac{dn}{dT}\right) \Delta T+\sigma_{\textup{FCD}}\Delta N \right) \label{eq_2.4}\\
\gamma & = 2\gamma_e+\gamma_i+\eta_{\textup{FCA}}\Delta N+\eta_{\textup{TPA}}|a|^2\label{eq_2.5}
\end{align}
\noindent where the meaning and the values of the various symbols are given in Table \ref{tab_2} of Appendix A. From Eq.(\ref{eq_2.1}), the power $P_d$ at the Drop port is obtained by $P_d = \gamma_eU_{\textup{int}}$ and $P_{\textup{abs}} = 2\gamma U_{\textup{int}}$.\\
\noindent As a first step, we replaced the constant free carrier lifetime $\tau_{\textup{fc}}$ in Eq.(\ref{eq_2.2}) with the expression in Eq.(\ref{eq_1}). We then numerically integrated Eqs.(\ref{eq_2.1}-\ref{eq_2.3}) by changing $\tau_{\textup{th}}$ until the simulated SP period matches the one of the experiment. Arbitrarily, we used as a reference the experimental SP trace for $P = 7\,\textrm{dBm}$ and $\Delta\nu = -10\,\textrm{GHz}$. The best value of $\tau_{\textup{th}}$ is found by repetitively minimizing the least squares error between the experimental and the simulated Drop signal by using a Particle Swarm Optimizer (PSO) \cite{PSO}. The best match is found with $\tau_{\textup{th}} = 270(10)\,\textrm{ns}$. Remarkably, we had to increase the FCD coefficient up to $\sigma_{\textup{FCD}} = -9.5\times 10^{-27}\,\textrm{m}^3$ in order to observe SP, which is $\sim 2-3$ times higher than the values reported in literature \cite{lin2007nonlinear}. Using the set of parameters listed Table \ref{tab_2} of Appendix A, we generated the stability and the SP frequency maps shown in Fig.\ref{fig_4}(a,b). A comparison between some simulated and experimental traces is shown in Fig.\ref{fig_4}(c) (labelled as "Model 1").
\noindent Clearly, the overlap (fraction of matching points) with the maps in Fig.\ref{fig_2}(a) is very poor. In contrast with the experiment, the SP region is totally confined at negative detunings, and  the frequency trend in Fig.\ref{fig_4}(b) does not follow the one in Fig.\ref{fig_2}(b). Moreover, as emerges from Fig.\ref{fig_4}(c), the predicted duty cycle is almost half the one which we measured. We also performed more refined analysis, in which, for each pair of $(\sigma_{\textup{FCD}},\tau_{\textup{th}})$, we generated the same maps in Fig.\ref{fig_4}(a,b), and we evaluated the overlap with the ones in Fig.\ref{fig_2}(a,b), but we never obtained overlaps higher than $60\%$. According to \cite{van2012simplified}, the SP region shifts towards positive detunings if we increase the relative weight $q$ between FCD and TOE. Using the same notation of\cite{van2012simplified}, this scales as $q\propto\frac{c_p}{\tau_{\textup{th}}\left( \frac{dn}{dT}\right)}\sqrt{\sigma_{\textup{FCD}}\tau_{\textup{fc}}}$, where $\left( \frac{dn}{dT}\right)$ is the thermo-optic coefficient and $c_p$ the specific heat. Since $\tau_{\textup{fc}}$ has been measured, and the continuity equation governing free carrier evolution is well established \cite{sze2006physics}, we focused our attention to the temperature equation. We let the specific heat and the thermal relaxation time to be free parameters, and we independently fit each experimental trace in the plane $(P,\Delta\nu)$. This produced the values of $c_p(P,\Delta\nu)$ and $\tau_{\textup{th}}(P,\Delta\nu)$ shown in Fig.\ref{fig_4}(d), while the related temporal traces are shown in Fig.\ref{fig_4}(c) (labelled as "Model 2"). Several considerations now arise, with the most striking being the fact that, on average, the specific heat should be $\sim 7-10$ times higher than its tabulated reference value $c_{\textup{p,ref}}=700\,\textrm{J}/\textrm{(kgK)}$ . Furthermore, $\tau_{\textup{th}}$ lies between $350-650\,\textrm{ns}$, and as $c_p$, shows a clear correlation with the SP frequency in Fig.\ref{fig_2}(b). In particular, the longer the SP period, the higher is the value of $c_p$ and $\tau_{\textup{th}}$ required to fit the  data. This is shown in Fig.\ref{fig_4}(d), which evidences a linear correlation between the variables, indicating that the $q$ parameter is constant in the plane $(P,\Delta\nu)$. Using the average values $\tau_{\textup{th}} = 450$ and $c_p = 8c_{\textup{p,ref}}$, we obtain $q\sim 5q_0$, where $q_0$ is the value of $q$ generating the maps in Fig.\ref{fig_4}(a,b). This confirms that the original model was underestimating the role of free carriers on the overall resonance shift. The agreement with the experimental traces in Fig.\ref{fig_4}(c) is excellent, but it is reached with the nonphysical assumption that $c_p$ should be almost one order of magnitude larger than the one reported in literature. Similarly, the thermal decay constant is $\sim 3.5$ times higher than the ones reported in \cite{mancinelli2014chaotic,zhang2014experimental}, despite the MR geometry and the material platform are almost the same. This suggests that they may be artifacts which fix the inadequacy of the temperature equation. 
\section{The temperature equation: FEM vs ODE predictions}
\label{section_5}
\begin{figure}[h!]
\centering\includegraphics[scale = 0.6]{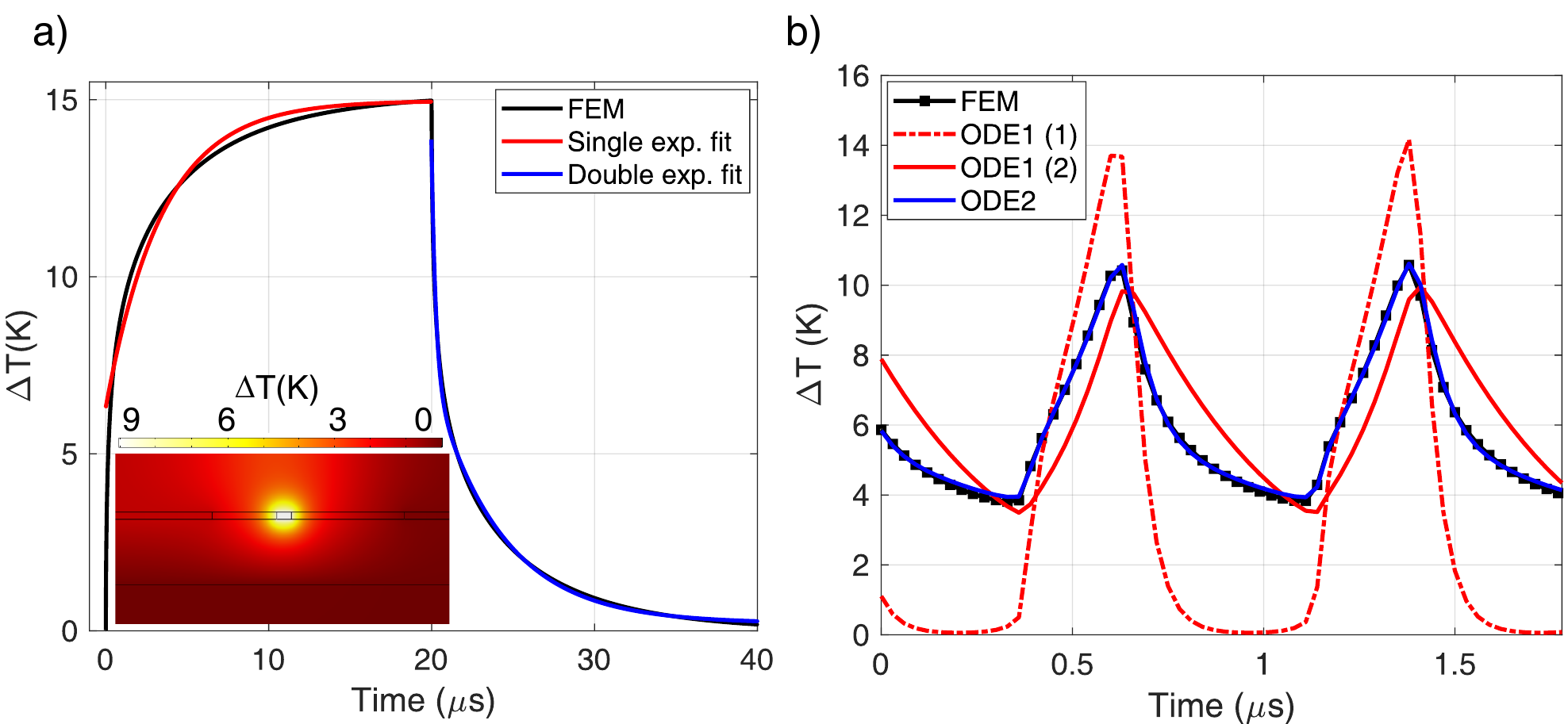}
\caption{(a) Temperature evolution $\Delta T$ (black) of the MR when a constant power of $2\,\textrm{mW}$ is applied through the whole cavity from $0$ to $20\,\mu\textrm{s}$. After $20\,\mu\textrm{s}$, the power is switched off. The temperature is obtained by solving the heat transfer equation with a commercial FEM solver (COMSOL Multiphysics \cite{comsol}), and by spatially averaging the temperature profile inside the core of the waveguide. The inset shows a cross section of the instantaneous distribution of $\Delta T$ at $t\sim21\,\mu\textrm{s}$ (axisymmetric view). The red line is a single exponential fit of the raising edge of $\Delta T$. The blue line fits the falling edge with a double exponential. (b) Temperature variation of the MR predicted by the FEM solver (black squares) through some cycles of SP for $(P=8\,\textrm{dBm},\Delta\nu=0\,\textrm{GHz})$. The driving power $P_{\textup{abs}}$ is obtained by solving Eqs.(\ref{eq_2.1}-\ref{eq_2.3}) with $c_p = 6.9c_{\textup{p,ref}}$ and $\tau_{\textup{th}} = 417\,\textrm{ns}$. The red dashed line (ODE 1 (1)) is obtained by solving Eqs.(\ref{eq_2.1}-\ref{eq_2.3}) using the same driving power $P_{\textup{abs}}$, but by forcing $c_p=c_{\textup{p,ref}}$. The value of the thermal lifetime $\tau_{\textup{th}}=40\,\textrm{ns}$ minimizes the least square error with the FEM trace. The solid red line (ODE1 (2)) is obtained by solving Eqs.(\ref{eq_2.1}-\ref{eq_2.3}) with $c_p=c_{\textup{p,ref}}$ and $\tau_{\textup{th}} = 417\,\textrm{ns}$. The solid blue line (ODE2) is obtained by integrating Eq.(\ref{eq_3}) and by minimizing the least square error with the FEM trace. This yielded $(\kappa = 8.87\times10^{-4}\,\textrm{ns}^{-1} ,\tau_f = 55.8\,\textrm{ns} ,\tau_s = 1200\,\textrm{ns} ,\eta = 0.31)$. \label{fig_5}}
\end{figure}

The ODE for temperature in Eq.(\ref{eq_2.3}) is derived from the Netwon's law of cooling, which has a single exponential as autonomous solution \cite{bergman2011fundamentals}. However, we can use a FEM solver (COMSOL Multiphysics \cite{comsol}) to show that Newton's law does not model the temperature dynamics of our MR. As an example, in Fig.\ref{fig_5}(a) we report the average temperature of the MR when a heat source supplying a costant power of $2\,\textrm{mW}$ for $20\,\mu\textrm{s}$ is placed in the waveguide core. These conditions emulate what happens during SP, where the absorbed power origins from TPA and FCA. 

\noindent The inset in Fig.\ref{fig_5}(a) indicates that the temperature profile in the surroundings of the MR is not spatially uniform. Simulations reveal that the external temperature distribution is evolving in time, since heat progressively propagates toward the outer regions of the silica cladding and beneath the silicon substrate. The region which embeds the silicon waveguide does not act as a thermal bath at a constant temperature, which is one of the assumptions behind  Newton's law. As a direct consequence, as shown in Fig.\ref{fig_5}(a), the raising(falling) edge of $\Delta T$ is not a single exponential. At this point, one could fix the discrepancy by directly coupling Eq.(\ref{eq_2.1},\ref{eq_2.2}) to the FEM solver, and integrate the temperature profile inside the MR to lump the spatial variables, i.e., $\Delta T(x,y,z,t) \rightarrow \Delta T(t)$. This is certainly possible, but on one hand is time consuming, and on the other is not suited for a linear stability analysis. Alternatively, one can construct an ODE which mimics the temporal solution given by the FEM. The starting point is the observation that a double exponential decay of the form $\Delta T = C_s\exp^{-\frac{t}{\tau_s}}+C_f\exp^{-\frac{t}{\tau_f}}$ (solid blue line in Fig.\ref{fig_5}(a)) well matches the FEM solution. Here, we call $\tau_s\gg\tau_f$ respectively the slow and the fast thermal lifetimes, while $C_1$ and $C_2$ are constants to be determined from the initial conditions. The associated ODE is $\frac{d^2\Delta T}{dt^2}+\left( \frac{1}{\tau_{\textup{eff}}} \right)\frac{d\Delta T}{dt}+\left( \frac{1}{\tau_s\tau_f} \right) \Delta T = 0
$, where $\tau_{\textup{eff}}=\left(\frac{1}{\tau_s}+\frac{1}{\tau_f}\right)^{-1}$. Equation (\ref{eq_3}) is homogeneous, so we must provide the driving terms related to the absorbed power. Heuristically, we propose the form:
\begin{equation}
    \frac{d^2\Delta T}{dt^2}+\left( \frac{1}{\tau_{\textup{eff}}} \right) \frac{d\Delta T}{dt}+\left( \frac{1}{\tau_s\tau_f} \right) \Delta T = \frac{\kappa P_{\textup{abs}}}{mc_p}+\eta\frac{dP_{\textup{abs}}}{dt} \label{eq_3}
\end{equation}
which finds justification from the following considerations. At steady state, the solution predicted by Eq.(\ref{eq_3}) is $\Delta T = \frac{\kappa \tau_s\tau_fP_{\textup{abs}}}{mc_p}$, which is analogous to the solution of Eq.(\ref{eq_2.3}) if we set $\tau_f\tau_s\kappa = \tau_{\textup{th}}$. As shown in Fig.\ref{fig_5}(a), $\frac{d\Delta T}{dt}$ must be discontinuous at time $t_0$, that is when  $P_{\textup{abs}}$ instantaneously drops from the ON to the OFF state. We can integrate Eq.(\ref{eq_3}) from $t_0^{-}=t_0-\xi$ to $t_0^{+}=t_0+\xi$ and let $\xi\rightarrow 0$ to obtain $\left. \frac{d\Delta T}{dt}\right|_{t_0^{+}}-\left. \frac{d\Delta T}{dt}\right|_{t_0^{-}} = \eta$, where we have used the fact that $\frac{dP_{\textup{abs}}}{dt}=\eta\delta(t-t_0)$. The parameters $(\kappa,\eta,\tau_s,\tau_f)$ are found by numerically integrating Eq.(\ref{eq_3}) and by simultaneously minimizing the least squares with the FEM temperature profile using a PSO. A realistic form of $P_{\textup{abs}}$ can be obtained by integrating Eqs.(\ref{eq_2.1}-\ref{eq_2.3}) using any of the values of $c_p$ and $\tau_{\textup{th}}$ shown in Fig.\ref{fig_4}(d). We used as reference $P = 8\,\textrm{dBm}$ and $\Delta\nu = 0\,\textrm{GHz}$. Figure \ref{fig_5}(b) shows a very good agreement between the temperature solution of the FEM and the one obtained by integrating Eq.(\ref{eq_3}).
\begin{figure}[h!]
\centering\includegraphics[scale = 0.66]{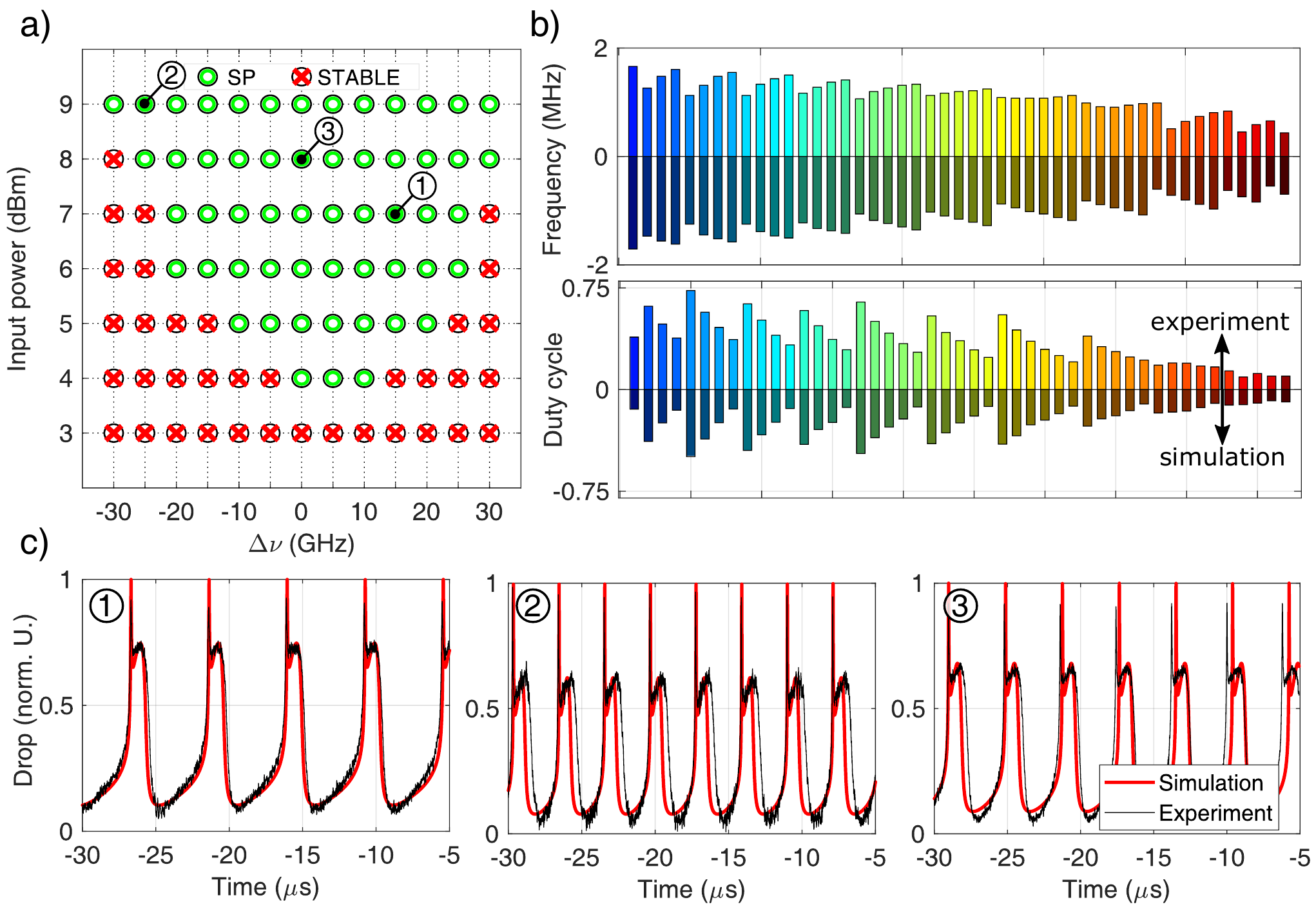}
\caption{(a) Simulated stability map which uses Eqs.(\ref{eq_2.1},\ref{eq_2.2}) for the energy and the carrier concentration and Eq.(\ref{eq_3}) for the temperature. The values of $(\kappa,\tau_f,\tau_s,\eta)$ are indicated in the main text. (b) Upper panel: comparison between the simulated (lower bars) and the experimental (upper bars) SP frequencies. The position in the plane $(P,\Delta\nu)$ is indicated by the same color code of Fig.\ref{fig_4}(d). Lower panel: same representation used in the upper panel, but relative to the duty cycle. This is defined as DC = $T_{\textup{th}}\nu_{\textup{SP}}$, where $T_{\textup{th}}$ is the total time in which the Drop trace lies above the threshold $0.45\cdot\textrm{(max[Drop]-min[Drop])}$, while $\nu_{\textup{sp}}$ is the SP frequency. (c) Examples of experimental (black) and simulated (red) Drop traces relative to the points (1), (2) and (3) in panel (a). \label{fig_6}}
\end{figure}
\noindent As a comparison, on the same plot we reported two curves, ODE1 (1) and ODE1 (2), which we obtain by integrating Eq.(\ref{eq_2.3}) and by setting respectively  $(c_{\textup{p,ref}},\tau_{\textup{th}}=40\,\textrm{ns})$ (the choice of $\tau_{\textup{th}}=40\,\textrm{ns}$ minimizes the error with the FEM trace) and $(c_p = 6.9c_{\textup{p,ref}},\tau_{\textup{th}} = 417\,\textrm{ns}$) (i.e., the pair of values in Fig.\ref{fig_4}(d)). The single order ODE is subjected to a trade-off between the decay time and the average temperature that is reached at stationary regime. The curve at $\tau_{\textup{th}}=40\,\textrm{ns}$ would benefit from a longer decay time, but this comes at the cost of an higher average temperature $T_{\textup{av}}$, since $T_{\textup{av}}\propto\frac{P_{\textup{abs}}\tau_{\textup{th}}}{c_p}$, which worsen the similarity with the FEM trace. This can be compensated by simultaneously increasing $c_p$ (solid red curve in Fig.\ref{fig_5}(b)), which is the origin of the anomalously high values reported in Fig.\ref{fig_4}(d). The tradeoff is removed in Eq.(\ref{eq_3}) by the inertial term $\left( \frac{1}{\tau_s\tau_f}\right)\Delta T$, which decouples the decay rate from the temperature at steady state. Equation (\ref{eq_3}) also explains why the values of $c_p$ and $\tau_{\textup{th}}$ are correlated to the SP period. As shown in Fig.\ref{fig_5}, the decay of $\Delta T$ has two characteristic lifetimes: one $\tau_f$ governing the initial stage, where the temperature is dropping fast, and another $\tau_s$ which takes over for $t>\tau_{f}$, creating a smooth decay tail. In the model of Eq.(\ref{eq_2.3}), $\Delta T$ decays as a single exponential, so ideally the match with Eq.(\ref{eq_3}) is lost as soon as $t>\tau_f$, that is, when the  slow exponential takes over. This can be partially recovered by increasing $\tau_{\textup{th}}$ as the overall decay time increases (i.e., the SP period), which compensates the absence of the slower exponential term. Concurrently, $c_p$ has to be increased to not alter the average temperature, which is why it gets linearly correlated with $\tau_{\textup{th}}$, as shown in Fig.\ref{fig_4}(d).\\
\noindent In a last step, we learned the parameters $(\kappa,\eta,\tau_s,\tau_f)$ using the experimental time response of the MR. This is accomplished by running a multi-objective PSO optimizer simultaneously minimizing the least square error between the MR output predicted by Eqs.(\ref{eq_2.1},\ref{eq_2.2},\ref{eq_3}) and three arbitrarily chosen experimental traces in the plane $(P,\Delta\nu)$. The result of the optimization yield $(\kappa = 7.45(5)\times10^{-4}\,\textrm{ns}^{-1} ,\tau_f = 71(2)\,\textrm{ns} ,\tau_s = 1030(20)\,\textrm{ns} ,\eta = 0.29(3))$, where the errors have been taken as the standard deviation over $10$ independent runs of the algorithm. We used this set to generate maps of stability, SP frequency and duty cycle, which we report in Fig.\ref{fig_6}(a,b). 
\noindent The improvement obtained by adopting the new model Eq.(\ref{eq_3}) is substantial. The overlap between the experimental and the simulated stability map raised from $41\%$ to $89\%$; the relative error on the SP frequency decreased from $13(7)\%$ to $7(8)\%$ and the one on the duty cycle from $60\%$ to $30(10)\%$. As can be appreciated from the time traces in Fig.\ref{fig_6}(c), all the characteristic features of SP discussed in Section \ref{section_2} are maintained by the new model given by Eq.(\ref{eq_3}). Worth to note, the improved agreement between the simulation and the experiments is obtained by solely correcting the temperature equation, with all the other parameters extracted from the experiment or taken from tabulated values (including $c_p=c_{\textup{p,ref}}$).

\section{Accuracy of the temperature equation}
\label{section_6}
\begin{figure}[h!]
\centering\includegraphics[scale = 0.5]{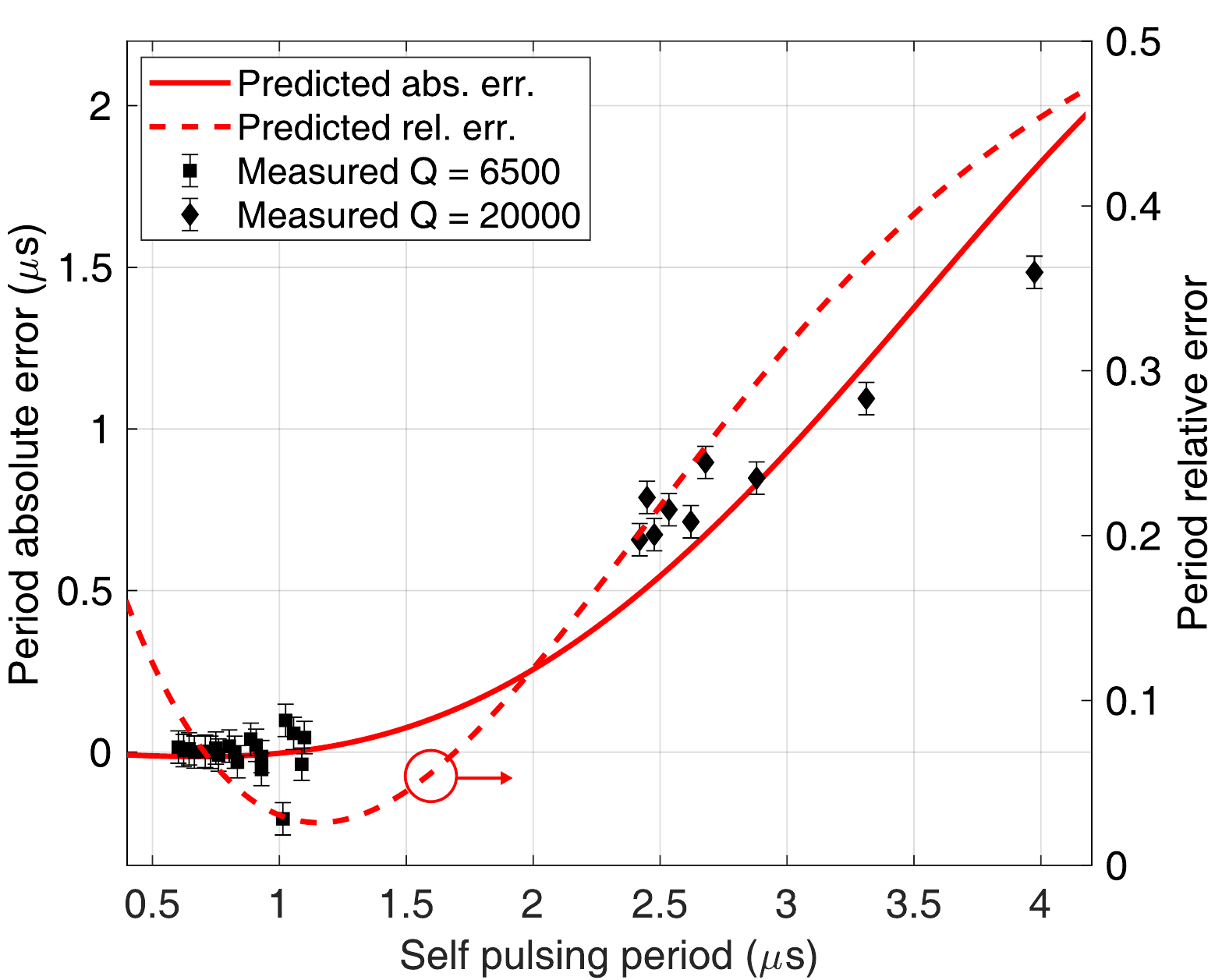}
\caption{Predicted absolute error (red line) on the SP period in case Eq.(\ref{eq_3}) is trained to reproduce the temperature evolution of the MR driven by a square wave with $2\,\textrm{mW}$ amplitude, $33\%$ duty-cycle and $1\,\mu\textrm{s}$ period. The relative error is shown with a red dashed line. Black squares indicate the error between the simulated and the experimental period for a MR of $Q=6500$. Black diamonds are relative to a MR of $Q=20000$, which naturally exhibits longer SP periods. Only the experimental data points associated to a duty-cycle in the range $[20-30]\%$ are shown for a fair comparison with the simulation. \label{fig_7}} 
\end{figure}

Equation (\ref{eq_3}) fixes the discrepancies between the model and the experiment by introducing an inertial term proportional to $\Delta T$ to increase the relative weight of FCD with respect to TOE . However, Eq.(\ref{eq_3}) is not isomorphic to the heat transfer equation (where we implicitly assumed that the spatial coordinates have been dropped by integrating the temperature profile over the MR volume). In this sense, it is an approximation whose range of validity has to be quantified. A natural question is whether or not it is possible to train the model on a specific SP frequency range and use it to predict the SP behaviour of a MR having different oscillation period. We addressed this question by training Eq.(\ref{eq_3}) to reproduce the solution of the FEM solver when the heat source is a square wave with $33\%$ duty-cycle, $2\,\textrm{mW}$ of amplitude, $100\%$ extinction and $T_{\textup{ref}} = 1\,\mu\textrm{s}$ of period. These conditions well approximate the temporal profile of $P_{\textup{abs}}$ during the SP regime of a MR. We then changed the period of the square wave to some other value $T_{\textup{sp}} \neq T_{\textup{ref}}$, and used the FEM solver to evaluate the maximum temperature excursion $\Delta T_{\textup{max}}$ for each SP cycle (initial transient effects are discarded in this analysis). For example, in Fig.\ref{fig_5}(b), $\Delta T_{\textup{max}} \sim 6\,\textrm{K}$ for $T_{\textup{sp}} = 0.8\,\mu\textrm{s}$. Note that $\Delta T_{\textup{max}}$ is a function of $T_{\textup{sp}}$.  We assume that the temperature variations predicted by FEM represent what occurs in a MR with a SP period of $T_{\textup{sp}}$. Using the same driving power $P_{\textup{abs}}$ in Eq.(\ref{eq_3}), we obtained the approximated ODE solution, from which we evaluated the time $T_{\textup{ODE}}$ required to reach the same temperature variation $\Delta T_{\textup{max}}$. The error on the predicted SP period by Eq.(\ref{eq_3}) is then determined by $T_{\textup{sp}}-T_{\textup{ODE}}$. This error is a function of the SP period and is shown as a solid red line in Fig.\ref{fig_7}. 
This approach is then experimentally validated. We measured the SP period $T_{\textup{exp}}$ of a MR with an identical perimeter as the one described in Section \ref{section_1}, but with $Q=20000$, and evaluated the difference with the SP period $T_{\textup{sim}}$ predicted by the full model given by Eqs.(\ref{eq_2.1},\ref{eq_2.2},\ref{eq_3}). To this purpose, we used the set of parameters 
$(\kappa,\eta,\tau_s,\tau_f)$ learned from the fits in Fig.\ref{fig_6}(c), which are related to a MR with $Q=6500$. The errors on the SP period $T_{\textup{exp}}-T_{\textup{sim}}$ as a function of $T_{\textup{sp}}$ are shown in Fig.\ref{fig_7} (black diamonds). As a comparison, we also included the errors on the SP period $T_{\textup{exp}}-T_{\textup{sim}}$ relative to the MR with $Q=6500$ (black squares).  As can be seen, the errors of the model with respect to the experiment or to the full FEM simulation are similar, which allows us to evaluate the reliability of the approximated temperature model. In general, Eq.(\ref{eq_3}) tends to underestimate the SP period, with an error that super-linearly increases with the latter. As expected, the relative error is at minimum where the model has been trained (i.e., $1\,\mu\textrm{s}$). If we set the desired accuracy threshold to $10\%$, the model loses accuracy as long as the period exceeds $2\,\mu\textrm{s}$ or gets lower than $500\,\textrm{ns}$. This interval is sufficiently large to accommodate all the SP frequencies reported in Fig.\ref{fig_2}(b).

\section{General remarks on the different modeling approaches}
\label{section_7}
In the previous section, we have seen that the model trained on a given MR does not generalize well to other MR with SP periods that are $\sim 2-3$ times higher(lower). It is then expected that different MR have to be modeled by different sets $(\kappa,\eta,\tau_s,\tau_f)$. This holds also for MR which share the same geometry but have different Q. This is of little use for the engineering of such dynamical systems, where we would like to know in advance their performances before fabrication and testing. An universal modeling of these systems can only be obtained if Eqs.(\ref{eq_2.1},\ref{eq_2.2}) are coupled to the FEM solution of the temperature equation. In order to reduce the computational effort, a possible strategy could be to use the FEM solver to obtain accurate solutions for only very few points in the phase space $(P,\Delta\nu)$, which are then used to train Eq.(\ref{eq_3}) to predict the time response for all the other points. We expect that Newton's law will nevertheless produce accurate solutions for SP periods of the order of $\tau_f$, since the slow thermal lifetime will appear as a simple constant background. However, as shown in Fig.\ref{fig_5}(b), this will not fix the need of increasing $c_p$ to reproduce the correct temperature excursion. In principle, we could also leave $c_p=c_{\textup{p,ref}}$, but we will have to artificially increase $\tau_{\textup{fc}}$ and/or $\sigma_{\textup{FCD}}$ in order to restore the relative weight between FCD and TOE. Similarly, a single exponential solution works well if the driving input power is changing much faster than $\tau_f$, as it happens for OOK signals transmitted at Gpbs data rates \cite{de2019power,zheng2012enhanced}. Indeed, these are effectively seen as CW signals from the point of view of the temperature evolution, so $\tau_{\textup{th}}$ can be tailored to match the steady state temperature observed in the experiment or predicted by the FEM simulation. \\
There could be also some cases for which the single exponential solution can be safely used without any sort of corrections and for every SP period, which are the ones where Newton's law of cooling holds. An example could be an air cladded silicon microdisk/microtoroid suspended on a silica pedestal, as the ones in \cite{johnson2006self,baker2012optical}. In this case, the substrate acts as a thermal bath with constant temperature, which justifies the validity of Newton's law. 

\section{Conclusions}
\label{section_8}
In this work, we refined the coupled equations describing the dynamics of silicon microresonators under thermal and free carrier nonlinearities. We showed that for MR buried in a silicon dioxide cladding, the Newton's law of cooling, which was traditionally used to model the waveguide temperature, can be inadequate. In particular, it leads to an overestimation of the thermo optic effect over free carrier dispersion. The immediate consequence, which we support by experimental evidences, is that these equations wrongly predicts the positions of the stability regions in the $(P,\Delta\nu)$ plane. Using inputs from FEM simulations, we proposed and implemented an alternative equation for the temperature evolution, which overcomes most of the limitations of Newton's law. The new equation introduces a slow and a fast thermal relaxation time, which can be both observed when, as in our case, the latter is comparable to the SP period. This decisively improved the agreement with the experiment of both the stability maps and the shape of the self pulsing waveforms. We envisage that FEM simulations could be used as inputs for training this ODE model for temperature, which simultaneously speeds up computation and allows to handle the linear stability analysis. As a last step, we discussed those conditions  where Newton's law can still lead to accurate predictions, but at the cost of introducing artifacts on the value of the specific heat, the carrier lifetime or the free carrier dispersion coefficient. We believe that this work completes the comprehension of a well established phenomena in silicon resonators, and can be helpful for the design and performance assessment of telecommunication devices or dynamical nodes for neuromorphic computing.
\section*{Appendix A: Values of the terms in the coupled equations}
Table \ref{tab_2} lists the definitions and the values of all the terms which appear in the set of coupled Eqs.(\ref{eq_2.1}-\ref{eq_2.3}), with the exception of $\tau_{\textup{fc}}$ and $\tau_{\textup{th}}$, that are defined in the main text. The resonance shift induced by free carriers is modeled as $\delta_{\textup{fc}} = -\frac{\sigma_{\textup{FCD}}\Delta N}{n_0}$ to produce the maps in Fig.\ref{fig_4}(a,b). The refined expression $\delta_{\textup{fc}} = -\frac{\sigma_{\textup{FCD,1}}\Delta N+\sigma_{\textup{FCD,2}}\Delta N^{0.8}}{n_0}$ \cite{lin2007nonlinear} is used to produce the maps in Fig.\ref{fig_6}(a,b).
\newpage
\begin{table}
\begin{tabular}[h!]{c|c|c|c}
Parameter & Description & Value/Definition & Source\tabularnewline
\hline 
$\omega_{0}$ & Resonance frequency & $2\pi\times192.88\,\textrm{THz}$ & Exp.\tabularnewline
$\gamma_{e}$ & Rate of extrinsic loss into the bus waveguide & $43\,\textrm{GHz}$ & Exp.\tabularnewline
$\gamma_{i}$ & Rate of intrinsic loss & $5.5\,\textrm{GHz}$ & Exp.\tabularnewline
$n_{0}$ & Refractive index of silicon & $3.485$ & \cite{johnson2006self} \tabularnewline
$\frac{dn}{d\Delta T}$ & Thermo optic coefficient of Silicon & $1.86\times10^{-4}\,\textrm{K}^{-1}$ & \cite{johnson2006self} \tabularnewline
$\sigma_{\textup{FCD}}$ & Lumped Free carrier dispersion coefficient & $-9.5\times10^{-27}\textrm{\,m}^{3}$ & Fit\tabularnewline
$\sigma_{\textup{FCD,1}}$ & Free carrier dispersion coefficient (electrons) & $-1.07\times10^{-27}\,\textrm{m}^{3}$ & Fit\tabularnewline
$\sigma_{\textup{FCD,2}}$ & Free carrier dispersion coefficient (holes) & $-1.63\times10^{-22}\,(\textrm{m}^{3})^{0.8}$ & Fit\tabularnewline
$\sigma_{\textup{FCA}}$ & Free carrier absorption coefficient & $1.5\times10^{-21}\,\textrm{m}^{2}$ & Fit\tabularnewline
$m$ & Resonator mass & $1.09\times10^{-14}\,\textrm{Kg}$ & Exp.\tabularnewline
$c_{\textup{p,ref}}$ & Specific heat of Silicon & $700\,\textrm{\ensuremath{\frac{\textrm{J}}{KgK}}}$ & \cite{johnson2006self} \tabularnewline
$g_{\textup{TPA}}$ & Free carrier generation rate per unit energy & $\frac{\Gamma_{\textup{fc}}c^{2}\beta_{\textup{TPA}}}{2n_{0}^{2}\hbar\omega_{0}V_{\textup{FCA}}^{2}}$ & \cite{zhang2013multibistability} \tabularnewline
$\eta_{\textup{TPA}}$ & Two photon absorption loss rate per unit energy & $\frac{\Gamma_{\textup{TPA}}c^{2}\beta_{TPA}}{2n_{0}^{2}V_{\textup{TPA}}}$ & \cite{zhang2013multibistability} \tabularnewline
$\Gamma_{\textup{FC}}$ & Free carrier confinement factor  & $0.999$ & FEM\tabularnewline
$\Gamma_{\textup{TPA}}$ & Two Photon Absorption confinement factor  & $0.992\,$ & FEM\tabularnewline
$V_{\textup{FCA}}$ & Free carrier effective volume & $4.87\times10^{-18}\,\textrm{m}^{3}$ & FEM\tabularnewline
$V_{\textup{TPA}}$ & Two Photon Absorption effective volume & $5.35\times10^{-18\,}\textrm{m}^{3}$ & FEM\tabularnewline
$\beta_{\textup{TPA}}$ & Two Photon absorption coefficient of Silicon & $8.1\times10^{-12}\textrm{\ensuremath{\frac{\textrm{m}}{W}}}$ & Fit \tabularnewline
$\eta_{\textup{FCA}}$ & Rate of loss due to free carrier absorption  & $\frac{c\sigma_{\textup{FCA}}}{n_{g}}$ & \cite{johnson2006self} \tabularnewline
$n_{g}$ & Group index of silicon waveguide & $4.4$ & Exp.\tabularnewline
$P_{\textup{abs}}$ & Total absorbed power & $2\gamma U_{\textup{int}}$ & \cite{johnson2006self}\tabularnewline
\end{tabular}
\caption{Definition, description and values of the parameters which appear in Eqs.(\ref{eq_2.1}-\ref{eq_2.3}) in the main text. \label{tab_2}}
\end{table}
\section*{Appendix B: Values of the terms in the Shockley Read Hall expression for $\tau_{\textup{fc}}$}
Trap assisted recombination at the Si/SiO$_2$ interface of the silicon waveguide is described using the well known Shockley Read Hall expression in Eq.(\ref{eq_1}) \cite{shockley1952statistics}. Here, it is assumed that the recombination is dominated by a single trap level, lying at energy $E_t$ inside the silicon energy gap $E_g$. The density of traps and the electron(hole) capture cross-section at the interface lumps into a phenomenological characteristic lifetime $\tau_n$($\tau_p$).
The terms $(a,b,\tau_0)$ appearing in Eqs.(\ref{eq_2.1}-\ref{eq_2.3}) have the following definitions:
\begin{equation}
\label{eq_4}
a =    \frac{\tau_p+\tau_n}{\tau_p(n_1+n_0)+\tau_n(p_1+p_0)} \quad\quad b  = \frac{1}{n_0+p_0} \quad \quad \tau_0  = \tau_p\frac{n_0+n_1}{n_0+p_0}+\tau_n\frac{p_0+p_1}{n_0+p_0}\end{equation}
in which all the above quantities are defined in Table \ref{tab_1}. 
\begin{table}[h]
\begin{tabular}{c|c|c|c}
Parameter & Definition & Value & Source\tabularnewline
\hline 
$\tau_{n}$ & Characteristic electron lifetime & $35\,\textrm{ns}$ & Fit\tabularnewline
$\tau_{p}$ & Characteristic hole lifetime & $22.5\,\textrm{ns}$ & Fit\tabularnewline
$n_{i}$ & Intrinsic equilibrium carrier concentration & $5\times10^{11}\,\textrm{cm}^{-3}$ & \cite{sze2006physics} \tabularnewline
$p_{0}$ & Hole concentration at equilibrium & $10^{15}\,\textrm{cm}^{-3}$ & Exp.\tabularnewline
$n_{0}$ & Electron concentration at equilibrium & $\frac{n_{i}^{2}}{p_{0}}$ & -\tabularnewline
$E_{g}$ & Silicon energy gap & $1.12\,\textrm{eV}$ & \cite{aldaya2017nonlinear}\tabularnewline
$E_{i}$ & Intrinsic Fermi level & $0.56\,\textrm{eV}$ & \cite{aldaya2017nonlinear}\tabularnewline
$E_{t}$ & Trap energy level & $0.66\,\textrm{eV}$ & Fit\tabularnewline
\multirow{2}{*}{$n_{1}$} & \multirow{2}{*}{\parbox{6.8cm}{Number of electrons in the conduction band when the Fermi level coincides with the trap level}} & \multirow{2}{*}{$n_{i}e^{\frac{E_{t}-E_{i}}{k_{b}T}}$} & \multirow{2}{*}{Fit}\tabularnewline
 & & \\
\multirow{2}{*}{$p_{1}$} & \multirow{2}{*}{\parbox{6.8cm}{Number of holes in the valence band in case the Fermi level coincides
with the trap level}} & \multirow{2}{*}{$n_{i}e^{-\frac{E_{t}-E_{i}}{k_{b}T}}$} & \multirow{2}{*}{Fit}\tabularnewline
& & \\
$T$ & Equilibrium temperature & $293.15\,\textrm{K}$ & Exp. \tabularnewline
$k_{b}$ & Boltzmann's constant & $8.61\times10^{-5}\,\textrm{eVK}^{-1}$ & \cite{sze2006physics} \tabularnewline
\end{tabular}

\caption{Definition, description and values of the parameters which appear in Eq.(\ref{eq_4}). \label{tab_1}}

\end{table}
\section*{Funding}
This project has received funding from the European Research Council (ERC) under the European Union’s Horizon 2020 research and innovation programme (grant agreement No 788793, BACKUP), and from the MIUR under the project PRIN PELM (20177 PSCKT).

\end{document}